\documentclass[prl,letterpaper,twocolumn,showpacs]{revtex4-1} 
\usepackage{times,xspace} 
\usepackage{amsbsy,amssymb,amsmath,bm,bbold} 
\usepackage{graphicx,color,epsfig,rotate} 
\usepackage{fancyhdr} 
\usepackage{epstopdf}
\usepackage{float}
\usepackage[colorlinks=true,linkcolor=blue,citecolor=blue,urlcolor=blue]{hyperref}

\def\bbbc{{\mathchoice {\setbox0=\hbox{$\displaystyle\rm C$}\hbox{\hbox 
to0pt{\kern0.4\wd0\vrule height0.9\ht0\hss}\box0}} 
{\setbox0=\hbox{$\textstyle\rm C$}\hbox{\hbox 
to0pt{\kern0.4\wd0\vrule height0.9\ht0\hss}\box0}} 
{\setbox0=\hbox{$\scriptstyle\rm C$}\hbox{\hbox 
to0pt{\kern0.4\wd0\vrule height0.9\ht0\hss}\box0}} 
{\setbox0=\hbox{$\scriptscriptstyle\rm C$}\hbox{\hbox 
to0pt{\kern0.4\wd0\vrule height0.9\ht0\hss}\box0}}}}

\begin{document} 
\title{Exceptional Point and Toward Mode Selective Optical Isolation}
\author{Arnab Laha,$^1$ Sibnath Dey,$^1$ Harsh K. Gandhi,$^1$ Abhijit Biswas,$^2$ and Somnath Ghosh$^{1,}$}
\email{somiit@rediffmail.com}
\affiliation{$^1$Department of Physics, Indian Institute of Technology Jodhpur, Rajasthan-342037, India\\
	$^2$Institute of Radiophysics and Electronics, University of Calcutta, Kolkata-700009, India}

\begin{abstract} 

Dynamical encirclement of an Exceptional Point (EP) and corresponding time-asymmetric mode evolution properties due to breakdown in adiabatic theorem have been a key to range of exotic physical effects in various open atomic, molecular and optical systems. Here, exploiting a gain-loss assisted dual-mode optical waveguide that hosts a dynamical EP-encirclement scheme, we have explored enhanced nonreciprocal effect in the dynamics of light with onset of saturable nonlinearity in the optical medium. We propose a prototype waveguide-based isolation scheme with judicious tuning of nonlinearity level where one can pass only a chosen mode in any of the desired directions as per device requirement. The deliberate presence of EP enormously enhances the nonreciprocal transmission contrast even up to 40 dB over the proposed device length with a scope of further scalability. This exclusive topologically robust mode selective all-optical isolation scheme will certainly offer opportunities in integrated photonic circuits for efficient coupling operation from external sources and improve device performances.
\end{abstract} 
 
 
\maketitle %

Nonreciprocal light propagation associated with breaking Lorentz's reciprocity in a nonlinear medium is the fundamental requirement to realize optical isolation; where time-reversal symmetry is not well-maintained. Isolators are indispensable in almost all optical systems, for example, protection of a high-power laser from back reflection, reduction of multipath interference in communication system and optical signal-processing (for recent review see ref. \cite{Caloz18}). To this end, the nonreciprocal transmission has commonly been achieved via magneto-optical Faraday rotation effect \cite{Shoji14}, however, on chip-scale integrated photonics, it remains elusive due to unavailability of necessary materials to achieve sufficient Faraday rotation effect. Thus, there is an usual challenge in production of all-optical isolators on chip-scale device footprint. Recently, there have been investigations to explore Exceptional Points (EPs) to achieve extraordinary nonreciprocal effects \cite{Thomas16,Choi17,Wang18}. Here, exploiting the chiral behavior of an EP, we look into a step-forward approach to realize a waveguide-based, mode-selective, topologically robust optical isolator. 


EPs are the topological branch point singularities that appear in the parameter space (at least 2D) of the usually dissipative non-Hermitian systems where the coupled eigenvalues and the corresponding eigenstates of the underlying Hamiltonian simultaneously coalesce \cite{Heiss00}. The unconventional physical effects in the vicinity of EPs \cite{Miri19} have attracted revolutionized attention with respect to a wide range of astonishing technological aspects that include unidirectional light propagation \cite{Huang17,Laha19}, asymmetric mode conversion/switching \cite{Laha19,Ghosh16,Doppler16}, lasing and anti-lasing \cite{Wong16}, ultra-sensitive optical sensing \cite{Wiersig16}, etc. Particularly, the chiral topological nature of an EP \cite{Dembowski01} and associated geometric phase behavior \cite{Mailybaev05} have been seen in the adiabatic state-exchange mechanism which is essentially reciprocal in nature. Here, a sufficiently slow parametric evolution along a closed loop around an EP allows the continuous swapping between the interacting modes/states \cite{Laha19,Ghosh16}. Now, if we consider the dynamical EP-encirclement scheme i.e. time (or analogous length-scale) dependent parametric variation then the contrast between the effect of EP and the standard adiabatic theorem enables nonadiabatic evolution of one of the two interacting eigenstates \cite{Gilary13,Milburn15} which leads to an asymmetric mode conversion phenomenon; where depending on the direction of rotation a specific eigenstate dominates at the end of encirclement process \cite{Ghosh16,Doppler16}.

Owing to the precise control of an EP on time-asymmetric modal dynamics, in this letter, we exploit this reciprocal light propagation phenomena to achieve huge nonreciprocity in a gain-loss assisted dual-mode optical waveguide \cite{Ghosh16}. This is made feasible by making the corresponding scattering matrix asymmetric with onset of suitable nonlinearity in the optical medium. Here, we propose an isolation scheme based on two different four-port prototypes which are topologically robust in the same design of the waveguide, however, having different amounts of nonlinearity. The designed waveguide supports a parameter space with longitudinal variation of gain-loss profile along the propagation direction that encircles the EP dynamically. Appreciating the associated EP-aided asymmetric mode conversion scheme, two proposed prototype isolators deliver two different specific modes in different directions based on the amount of nonlinearity. Here, the allowed mode is selected by the non-adiabatic corrections around EP in the respective direction.

\begin{figure}[t]
	\centering
	\includegraphics[width=8.5cm]{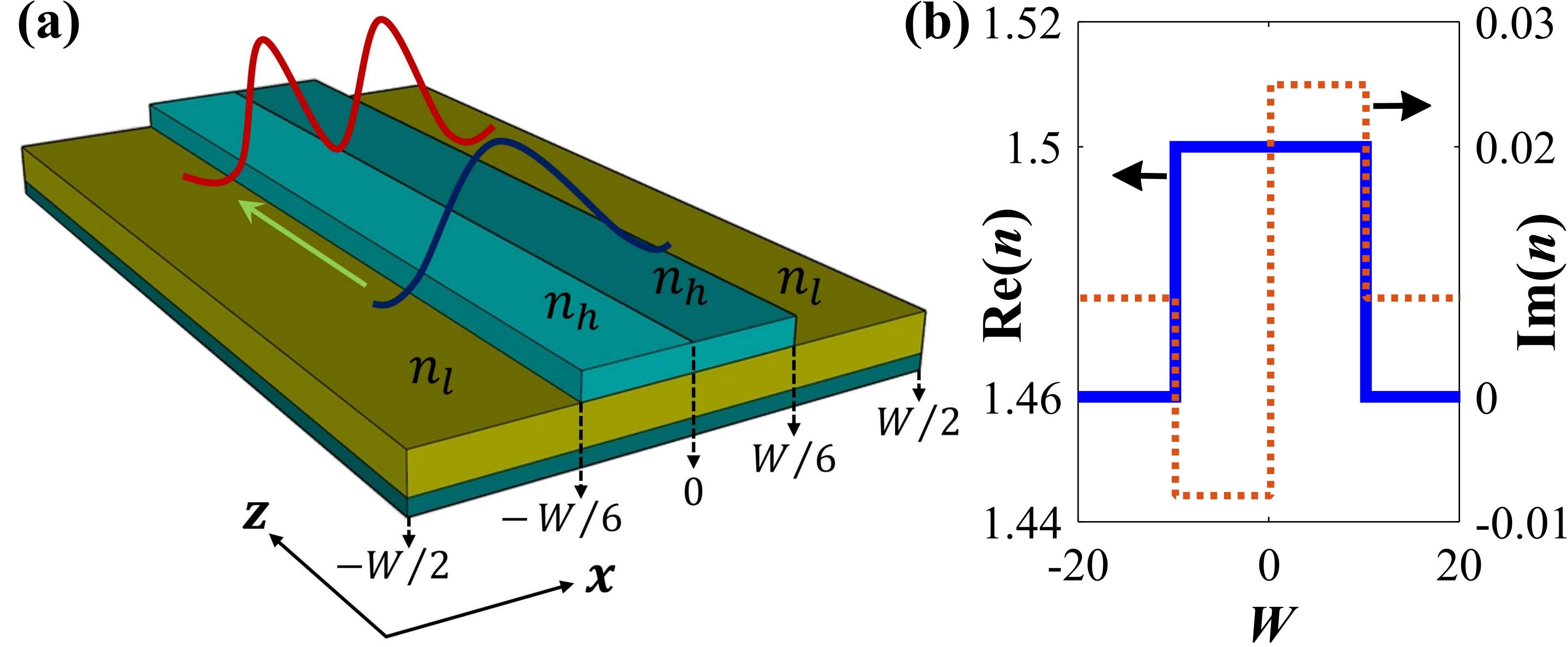}
	\caption{\textbf{(a)} Schematic of the proposed specialty optical waveguide. $x$ and $z$ represent the transverse and propagation directions respectively. \textbf{(b)} Transverse profile of $n(x)$: Blue solid line shows the $\text{Re}(n)$, and the brown dotted line shows the $\text{Im}(n)$ for the specific parameter set ($\gamma=0.008,\,\tau=3.161$) at EP.}
	\label{structure}
\end{figure}

We consider a customized step-index planar optical waveguide \cite{Ghosh16} that occupies the region $-W/2\le x\le W/2$ as shown in Fig. \ref{structure}(a). Normalizing the operating frequency $\omega=1$, we set the total width $W=20\lambda/\pi=40$ in a dimensionless unit (feasibly, one can choose $\mu m$). The fixed real refractive indices of the core and cladding have been chosen as $n_h=1.5$ and $n_l=1.46$, respectively. For these specified operating parameters, the waveguide supports the fundamental mode ($\rm{LP_{01}}$, say, $\psi_0$) and the first-higher-order mode ($\rm{LP_{11}}$, say, $\psi_1$). 

Now, beyond PT-symmetry, we introduce non-Hermiticity in the designed waveguide by integrating a transverse distribution of unbalanced gain-loss profile that can be realized with the patterned imaginary part of the refractive index. Along the longitudinal direction, the amount of non-Hermiticity is modulated by independent tunabilities of two control parameters viz. the gain coefficient: $\gamma(z)$ and the loss-to-gain ratio: $\tau(z)$. Thus, the overall transverse refractive index profile for a specific cross-section of the waveguide can be written as
\begin{equation}
n(x)=\left\{ 
\begin{array}{ll}
\vspace{0.1cm}
n_h-i\gamma, &\quad -W/6\le x\le 0\\
\vspace{0.1cm}
n_h+i\tau\gamma, &\quad 0\le x\le W/6\\
\vspace{0.1cm}
n_l+i\gamma, &\quad W/6\le |x|\le W/2.
\end{array}
\right.
\label{nx} 
\end{equation}
The profiles of $\text{Re}(n)$ and $\text{Im}(n)$ (for a fixed $\gamma$ and $\tau$) are shown in Fig. \ref{structure}(b). Now, the propagation constants ($\beta$-values) of the supported modes are computed from the scalar modal equation $[\partial_x^2+n^2(x)\omega^2-\beta^2]\psi(x)=0$, using $n(x)$ as given by Eq. \ref{nx}.

\begin{figure}[t]
	\centering
	\includegraphics[width=8.5cm]{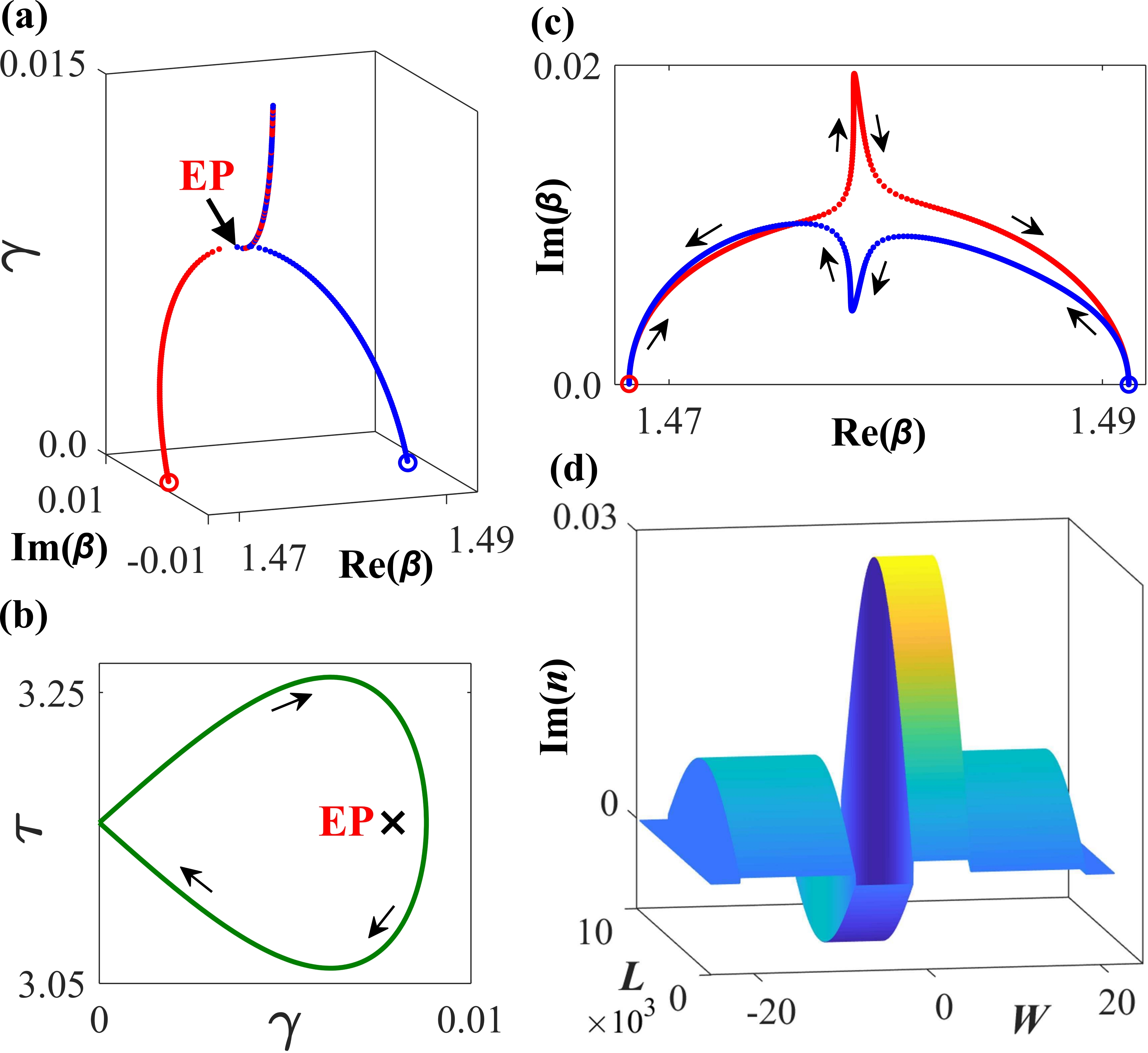}
	\caption{\textbf{(a)} Encounter of an EP: Trajectories of the complex $\beta_0$ (dotted blue curve) and $\beta_1$ (dotted red curve) for a chosen $\tau=3.17$ with an increase in $\gamma$, showing coalescence at EP near $\gamma=0.008$. The circular markers of the respective colors represent their initial position at $\gamma=0$. \textbf{(b)} EP-encirclement: Clockwise variation of $\gamma$ and $\tau$ around the EP. \textbf{(c)} $\beta$-switching: Trajectories of complex $\beta$-values following the parametric loop shown in (b). Arrows indicate the direction of progression. \textbf{(d)} Parameter space mapping on the waveguide: Variation of $\text{Im}(n)$ along both $x$ and $z$ directions.}
	\label{EP}
\end{figure}

With the introduction of gain-loss, $\psi_0$ and $\psi_1$ are mutually coupled. Exploiting the concept of transition between topologically dissimilar avoided resonance crossings (ARCs), we encounter an EP where the corresponding complex $\beta$-values coalesce \cite{Laha19,Ghosh16}. We study the topological dynamics of $\beta_0$ and $\beta_1$ and corresponding ARC with crossing/anticrossing of their real and imaginary parts for different $\tau$ values, while $\gamma$ varies in a chosen range from 0 to 0.015. Judiciously examining several cases, we set a specific $\tau=3.17$ and track the dynamics of $\beta_0$ and $\beta_1$ in Fig. \ref{EP}(a). Here, it is evident that near $\gamma\simeq0.008$, they coalesce. Thus, we numerically detect the EP in the ($\gamma,\tau$)-plane at ($\gamma_{EP}=0.008,\,\tau_{EP}=3.161$). 

We execute an EP-encirclement process with consideration of a device implementation feasible loop in the ($\gamma,\tau$)-plane as given by the coupled equations $\gamma(\phi)=\gamma_{0}\sin(\phi/2)$ and $\tau(\phi)=\tau_{EP}+a\sin(\phi)$. Here, the characteristics parameters $a\,(\in(0,1])$ and $\phi\,(\in[0,2\pi])$ control the adiabaticity in variation of $\gamma$ and $\tau$ around EP. For $\gamma_0>\gamma_{EP}$, the parameter space encloses EP properly. $a>0$ confirms clockwise evolutions whereas $a<0$ allows anticlockwise evolutions. Fig. \ref{EP}(b) shows such an quasi-static encirclement scheme with $\gamma_0=0.009\,(>\gamma_{EP})$ and $a=0.1$. Following this specific contour in the ($\gamma,\tau$)-plane, we track the corresponding dynamics of $\beta_0$ and $\beta_1$ from their passive locations (where $\gamma=0$) in Fig. \ref{EP}(c). Here one complete encirclement around the EP allows the adiabatic permutation (mutual exchange in position) between $\beta_0$ and $\beta_1$ which revels the exact second order behavior of the identified EP.

\begin{figure}[t]
	\centering
	\includegraphics[width=\linewidth]{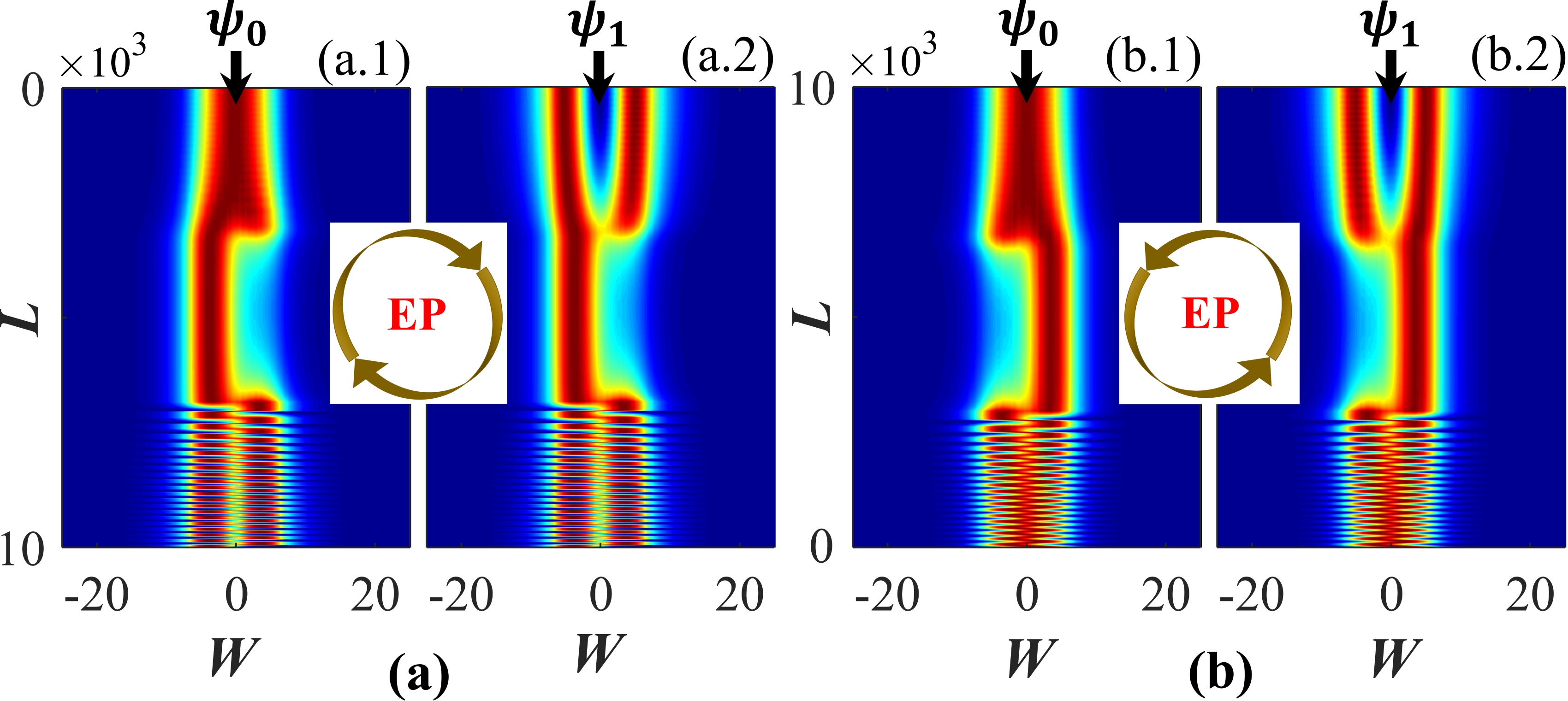}
	\caption{\textbf{(a)} Modal propagations from P1 to P2 during clockwise EP-encirclement; (a.1) adiabatic conversion from $\psi_0$ to $\psi_1$ and (a.2) nonadiabatic evolution of $\psi_1\,(\rightarrow\psi_1)$. \textbf{(b)} Modal propagations from P2 to P1 during anticlockwise EP-encirclement; (b.1) nonadiabatic evolution of $\psi_0\,(\rightarrow\psi_0)$ and (b.2) adiabatic conversion from $\psi_1$ to $\psi_0$. Due to proper visibility we re-normalize the modal intensities at each $z$ and hence the overall intensity variations are essentially scaled.} 
	\label{BP}
\end{figure}

Now, with the chosen parametric loop in the ($\gamma,\tau$)-plane as shown in Fig. \ref{EP}(b), we realize the dynamical EP-encirclement process by distributing the corresponding $\text{Im}(n)$ profile along length of the waveguide as shown in Fig. \ref{EP}(d). Under paraxial approximation, such a parameter space mapping should follow the time-dependent Schr{\"o}dinger equation with $z$ as the time axis. Denoting $L_0$ as the total operating length, the mapping equations can be written as
\begin{subequations}
	\begin{align}
	&\gamma(\phi)=\gamma_{0}\sin(\pi L_0/z);\quad\gamma_0>\gamma_{EP},\\
	&\tau(\phi)=\tau_{EP}+a\sin(2\pi L_0/z).
	\end{align}
	\label{device} 
\end{subequations}
We assign two ends of the waveguide as P1 at $z=0$ (i.e., $\phi=0$) and P2 at $z=L_0$ (i.e., $\phi=2\pi$). Here, one complete pass ($0\le z\le L_0$) along the length-axis describes an exact dynamical EP-encirclement scheme ($0\le\phi\le2\pi$) with simultaneous variation of $\gamma$ and $\tau$. Here, the clockwise and anticlockwise parametric evolutions around EP are simply achieved by changing the propagation directions. Now, dynamically encircling the EP with quasi-static gain-loss variation in the waveguide having total length $L_0=10^4$ (dimensionless unit), we exhibit the beam propagation of $\psi_0$ and $\psi_1$ in Fig. \ref{BP}. Once the variation of $\text{Im}(n)$ meets the adiabatic limit, the modal propagations can be simulated using split-step Fourier method using the modified scalar beam-propagation equation $2i\omega\partial_z\psi(x,z)=-[\partial_x^2+\Delta n^2(x,z)\omega^2]\psi(x,z)$ under paraxial approximation \cite{Agrawal12}. Here, $\Delta n^2(x,z)\equiv n^2(x,z)-n_l^2$.  

During dynamical EP-encirclement process, breaking of inversion symmetry occurs in the overall loss variation along the length-scale which results in adiabatic breakdown in the system-dynamics. Here, one of the interacting eigenstate that experiences comparably higher loss in comparison to its coupled counterpart behaves non-adiabatically. The average loss of the individual modes can be estimated by $\oint\{\text{Im}(\beta)/2\pi\}d\phi$, using adiabatic expectation of corresponding $\text{Im}(\beta)$ from Fig. \ref{EP}(c). Thus, for clockwise encirclement, $\psi_0$ and $\psi_1$ which have been individually excited from P1 ($z=0$) are essentially converted into $\psi_1$ at P2 ($z=L_0$) as can be seen in plots a.1 and a.2 in Fig. \ref{BP}(a). However, for any mode excited from P2 during anticlockwise encirclement, the waveguide yields $\psi_0$ at P1 as shown in b.1 and b.2 in Fig. \ref{BP}(b). The conversion efficiency of an input mode ($\psi^{\text{in}}$) has been calculated by its overlap integral with corresponding output mode ($\psi^{\text{op}}$) as \cite{Ghosh16}
\begin{equation}
C=\frac{\left|\int\left(\psi^{\text{op}}\times\psi^{\text{in}}\right) dx\right|^2}{\int\left|\psi^{\text{op}}\right|^2dx\int\left|\psi^{\text{in}}\right|^2dx}.
\label{C}
\end{equation} 
Here, during the clockwise rotation around the EP, we find $C_{\circlearrowright}$ as $92.82\%\,(\pm0.02\%)$ for the conversions $\{\psi_0,\psi_1\}\rightarrow\psi_1$; whereas during anti clockwise rotation, $C_{\circlearrowleft}=62.55\%\,(\pm0.04\%)$ for the conversions $\{\psi_0,\psi_1\}\rightarrow\psi_0$. 

Such bi-directional light transmissions in the optical waveguide can be realized by assigning a scattering matrix which is formulated as
\begin{equation}
[\psi^{\text{op}}]=[S]\,[\psi^{\text{in}}],\,\,\text{with}\,\, S=\left(\begin{array}{cccc}0&0&S_{13}&S_{14}\\0&0&S_{23}&S_{24}\\S_{31}&S_{32}&0&0\\S_{41}&S_{42}&0&0\end {array}\right).
\label{S}  
\end{equation}
Here, the zeros in top left and bottom right blocks of the $S$-matrix are chosen to neglect the all possible reflections. Top right block (say, $M_{tr}$) estimates the backward transmission as $T_B=|\max(M_{tr})|^2$ and bottom left block (say, $M_{bl}$) estimate the forward transmission as $T_F=|\max(M_{bl})|^2$. The nonzero matrix elements have been calculated using as $S_{ij}=\left\langle\psi^{\text{in}}_j\big|\psi^{\text{op}}_i\right\rangle$ with possible combinations of eigenfunctions at input ($\psi^{\text{in}}$) and output ($\psi^{\text{op}}$) of the waveguide.
\begin{figure}[t]
	\centering
	\includegraphics[width=6.5cm]{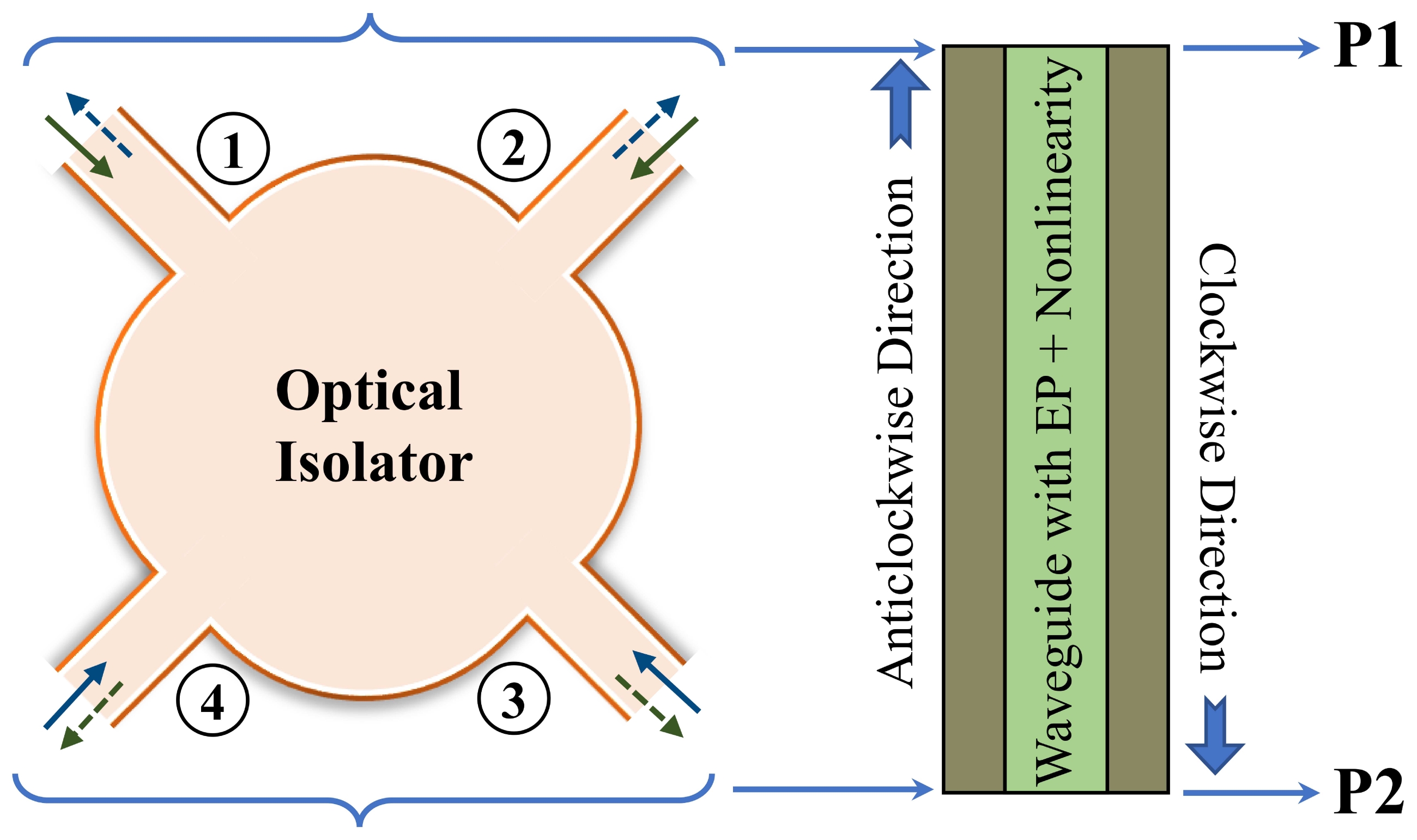}
	\caption{Schematic of a 4-port optical isolator on framework of the proposed waveguide with suitable nonlinearity. Ports (1,2) and (3,4) of the isolator are associated with waveguide ends P1 and P2, respectively.} 
	\label{isolator}
\end{figure}

In the linear interaction regime, the EP-aided chiral light transportation phenomena follow Lorentz's reciprocal theorem where the associated $S$-matrix must be symmetric ($S=S^T$). Now isolation is achieved when the waveguide allows only one-way-traffic, i.e., light is blocked in any one direction and allowed to pass in opposite direction. This implies that the nonreciprocity in light propagation demands breaking in Lorentz's reciprocity with an asymmetric $S$-matrix ($S\ne S^T$). Now, to break reciprocity, we introduce saturable nonlinearity, having the form $\Delta n_{\text{NL}}(x,z)=n_2|\psi|^2/\left[1+(|\psi|^2/I_s)\right]$, in the optical medium \cite{Srivastava11}. Here, $n_2$ is the nonlinear coefficient and $I_s$ is the saturating intensity. Unlike Kerr-nonlinearity, the chosen saturable nonlinearity has been incorporated to avoid unstable output while introducing nonreciprocity. 

Fig. \ref{isolator} describes such a schematic 4-port optical isolator in the framework of the designed dual-mode optical waveguide with saturable nonlinearity. The waveguide hosts the parameter space that enables to exhibit an EP-aided chirality driven mode-conversion phenomena as described in Fig. \ref{BP}. Here, we propose a prototype isolation scheme having two variants in the same operating configuration of the waveguide, while, having different amount of non-linearities (to distinguish them, say, WG1 and WG2). To distinguish the amount of nonlinearity, we normalize the factor $I_s$ and calibrate $n_2$; where the actual nonlinearity is quantified in the form of $\left(\Delta n_{\text{NL}}/\Delta n\right)\times100\%$ with $\Delta n=(n_h-n_l)$.
\begin{figure}[t]
	\centering
	\includegraphics[width=6.5cm]{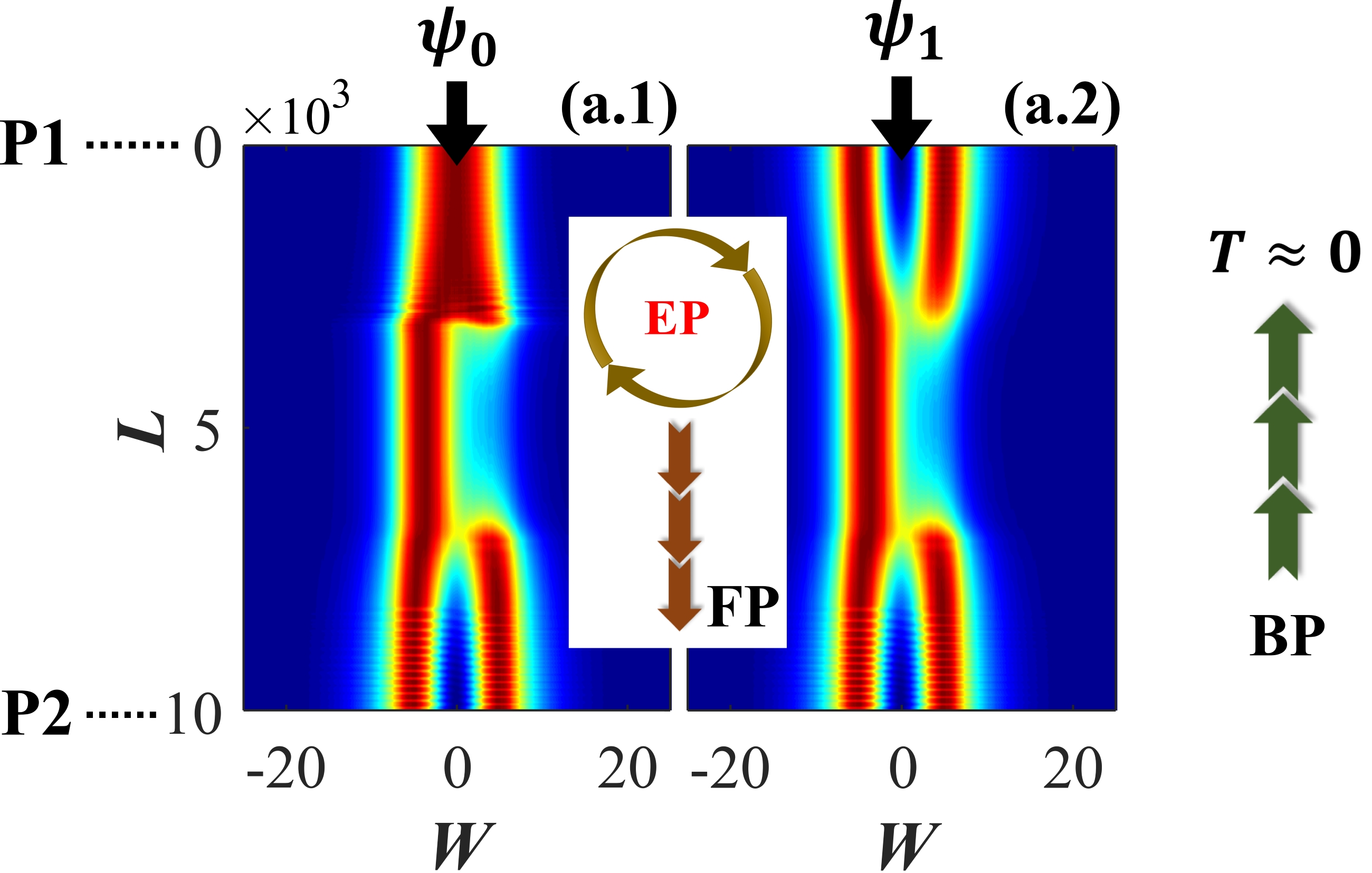}
	\caption{WG1: Prototype waveguide isolator with $3.75\%$ nonlinearity in the optical medium where one-way-traffic is active in the forward direction (from P1 to P2). Forward propagations of (a.1) $\psi_0$ and (a.2) $\psi_1$ from P1 to P2 for clockwise dynamical EP-encirclement scheme where both of them are converted to $\psi_1$ at P2. For backward propagation, $T\approx0$.}
	\label{p1}
\end{figure}

First, we introduce $3.75\%$ nonlinearity in the spatial index distribution of the waveguide and denote this prototype scheme as WG1. Now, to consider the clockwise dynamical EP-encirclement scheme we individually launch the coupled modes with unit magnitude from the P1-end. In Fig. \ref{p1} we study the propagations of $\psi_0$ (plot a.1) and $\psi_1$ (plot a.2). Here, along the forward direction, both $\psi_0$ and $\psi_1$ are almost fully transmitted through the waveguide. However, along this direction the EP-aided nonadiabatic corrections yield the dominating mode $\psi_1$ at P2 for both the cases. Meanwhile, when the modes are propagating in the backward direction i.e. for anticlockwise EP-encirclement process, both the modes that have been launched from P2-end of the waveguide are blocked from transmitting to P1-end. Here, due to breaking of inversion symmetry in overall gain-loss variation in presence of tailored nonlinearity along $z$-direction, major portion of the incoming intensities get attenuated, where overall transmission $T\approx0$. We calculate the nonzero elements of the asymmetric $S$-matrix and quantify the nonreciprocal effect in the prototype nonlinear waveguide WG1 in terms of isolation ratio using the expression $10\times\log_{10}(T_F/T_B)$; where maximum $\sim40.27$ dB isolation is accomplished.
\begin{figure}[b]
	\centering
	\includegraphics[width=6.5cm]{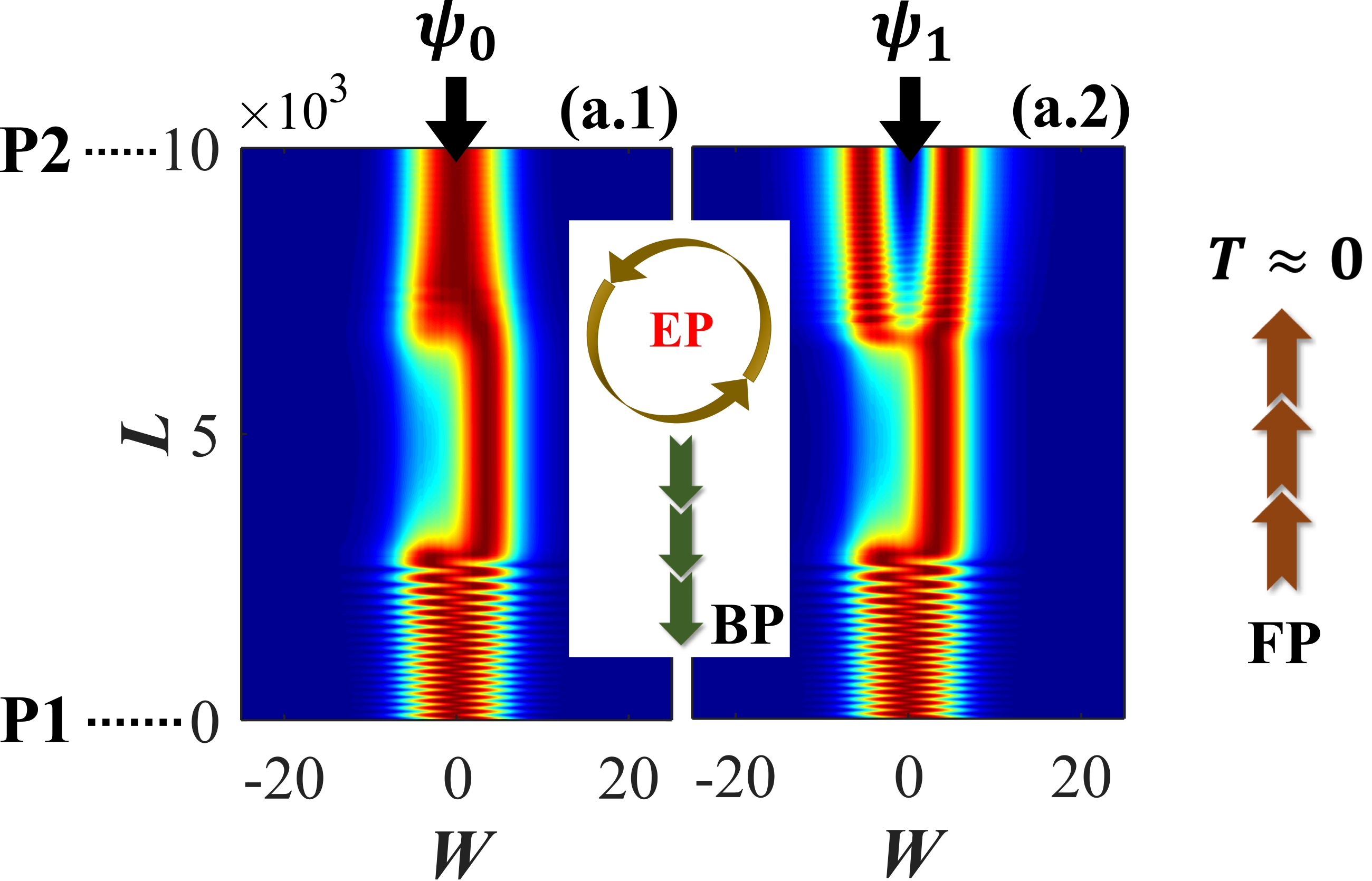}
	\caption{WG2: Prototype waveguide isolator with $8.75\%$ nonlinearity in the optical medium where one-way-traffic is active in the backward direction (from P2 to P1). Normalized backward propagations of (a.1) $\psi_0$ and (a.2) $\psi_1$ from P2 to P1 for anticlockwise dynamical EP-encirclement scheme where both of them are converted to $\psi_0$ at P1. Here, for forward propagation, $T\approx0$.} 
	\label{p2}
\end{figure}

Now, we increase the nonlinearity to $8.75\%$ in the optical medium of the waveguide having same spatial gain-loss distribution as described earlier (in Figs. \ref{BP} and \ref{p1}) and designate this prototype scheme as WG2. Here for clockwise EP-encirclement process i.e. when the modes are excited from the P1, there is no transmission ($T\approx0$) in the froward direction (from P1 to P2). Now, when we consider the anticlockwise EP-encirclement scheme with launching of $\psi_0$ and $\psi_1$ from P2, both of them are well-transmitted in the backward direction; and owing to nonadiabatic corrections in this direction, they are converted to the dominating mode $\psi_0$ at P1, as can be seen in plot a.1 and a.2 of Fig. \ref{p2}, respectively. Since, here the one-way-traffic is active from the backward propagation, the isolation ratio has been estimated using the expression $10\times\log_{10}(T_B/T_F)$; where we have achieved maximum $\sim21.55$ dB isolation.

Also, considering the longitudinal gain-loss variation in the waveguide for which EP is not properly enclosed and accordingly the modes are not converted, we have noted-down the isolation ratios for different amounts of optical nonlinearity up to $9\%$. Investigating such cases, we have achieved maximum $6.5$ dB (approximate) isolation. Thus, the presence of EP in parameter space results in giant increase of nonreciprocal effect (up to $\approx40$ dB isolation) for a comparably low nonlinearity level. Here, the presence of EP acts as an intrinsic tool to isolate a specifically selective mode. We have also calculated the mode conversion efficiencies for both the prototypes by using Eq. \ref{C}. In the prototype WG1, we estimate $C_{\circlearrowright}=98.13\%\,(\pm0.01\%)$ and for WG2 we obtain $C_{\circlearrowleft}=74.44\%\,(\pm0.01\%)$. Here, it should be noticeable that the conversion efficiencies increase in comparison with the cases when nonlinearity is not present in the optical medium. This aspect is also evident in Figs. \ref{p1} and \ref{p2}, where clarity of the converted outputs are reasonably enhanced.     

In summary, a topologically robust and versatile dual operation optical isolation scheme has been reported on the framework of a dual-mode optical waveguide operating at an EP. Based on the amount of nonlinearity, our designed nonreciprocal optical waveguide reverses its one-way-active traffic regulation, when we increase the amount of nonlinearity to a certain limit.  Here, owing to breaking inversion symmetry in gain-loss variation along $z$-axis, and corresponding EP-aided nonadiabatic corrections, our waveguide based two prototype isolators individually deliver different dominating modes at the opposite ports. Taking the advantage of giant nonreciprocal effect near the EP, we can set the desired nonlinearity level in the optical medium of the waveguide to isolate a selective mode in any of the required directions. Here, the presence of EP enormously increases the isolation ratio. Also, the presence of nonlinearity enhances the EP-aided mode conversion efficiencies which reflects the fact that both the proposed prototypes can efficiently be  used as one-way-mode-converter in integrated photonic devices. Aside from strong impact in fundamental physics, with suitable scalability and using state-of-the-art techniques, our scheme should open up an extensive platform to realize a new class of all-optical isolators in integrated device footprint. The predicted performance strongly encourages the experimental realization of the device.

\vspace{0.5cm}
A.L. and S.G acknowledge the financial support from the Science and Engineering research Board (SERB) [Grant No. ECR/2017/000491], Department of Science and Technology, Government of India. S.D. acknowledges the support from Ministry of Human Resource Development (MHRD), Government of India.

\end{document}